\documentclass[lettersize,journal]{IEEEtran}
\usepackage{amsmath,amsfonts}
\usepackage{algorithmic}
\usepackage{algorithm}
\usepackage{array}
\usepackage[caption=false,font=normalsize,labelfont=sf,textfont=sf]{subfig}
\usepackage{textcomp}
\usepackage{stfloats}
\usepackage{url}
\usepackage{verbatim}
\usepackage{graphicx}
\usepackage{cite}
\begin{document}

\title{Unveiling the Impact of Cognitive Distraction on Cyclists Psycho-behavioral Responses in an Immersive Virtual Environment}

\author{Xiang Guo, Arash Tavakoli, T. Donna Chen, and Arsalan Heydarian
\thanks{Corresponding Author: Dr. Arsalan Heydarian, ah6rx@virginia.edu}
}

\markboth{Journal of \LaTeX\ Class Files,~Vol.~14, No.~8, August~2021}%
{Shell \MakeLowercase{\textit{et al.}}: A Sample Article Using IEEEtran.cls for IEEE Journals}


\maketitle

\begin{abstract}
The National Highway Traffic Safety Administration reported that the number of bicyclist fatalities has increased by more than 35\% since 2010. One of the main reasons associated with cyclists' crashes is the adverse effect of high cognitive load due to distractions. However, very limited studies have evaluated the impact of secondary tasks on cognitive distraction during cycling. This study leverages an Immersive Virtual Environment (IVE) simulation environment to explore the effect of secondary tasks on cyclists' cognitive distraction through evaluating their behavioral and physiological responses. Specifically, by recruiting 75 participants, this study explores the effect of listening to music versus talking on the phone as a standardized secondary tasks on participants' behavior (i.e., speed, lane position, input power, head movement) as well as, physiological responses including participants' heart rate variability and skin conductance metrics. Our results show that  (1) listening to high-tempo music can lead to a significantly higher speed, a lower standard deviation of speed, and higher input power. Additionally, the trend is more significant for cyclists who had a strong habit of daily music listening ($>$ 4 hours/day).  In the high cognitive workload situation (simulated hands-free phone talking), cyclists had a lower speed with less input power and less head movement variation. Our results indicate that participants' HRV (HF, pnni-50) and EDA features (numbers of SCR peaks) are sensitive to cyclists' cognitive load changes in the IVE simulator.

\end{abstract}


\section{Introduction}
\subsection{Cyclist Crashes and Safety}
\IEEEPARstart{B}{icycle} users are expanding as an increasing number of cities are encouraging low-carbon transportation by investing in infrastructure to accommodate bicyclists \cite{pucher2010infrastructure}. This increasing trend hasn't been slowed down during the COVID-19 pandemic, and in fact it is reported that bicycling levels have significantly increased in many countries\cite{buehler2021covid}. However, bicyclist fatalities are increasing as the number of bicyclists road users is increasing over the last decade. The National Highway Traffic Safety Administration reports that the number of bicyclist fatalities has increased by more than 35\% since 2010 \cite{national2021fatality}. In addition to the increased number of cyclists on the road as mentioned above, several factors are believed to contribute to this alarming trend: lack of cycling infrastructures \cite{mulvaney2015cycling}, poor roadway designs \cite{guo2023psycho},  reckless driving and cycling behavior \cite{huemer2018influences,oldmeadow2019driver}, distraction \cite{useche2018distraction}, lack of public education \cite{pucher2008making}, and driver-centric vehicle design (e.g., smaller windows, wider pillars, larger headrests) \cite{koustanai2008statistical}. Among those factors, distraction has been identified as one of the main reasons for traffic accidents. In the US, nine percent of fatal crashes, 15 percent of injury crashes, and 15 percent of all police-reported motor vehicle traffic crashes in 2019 were reported as distraction-affected crashes. There were 566 vulnerable road users including pedestrians, pedal cyclists, and others killed in distraction-affected crashes \cite{national2019traffic}. The data for the fatalities and injuries are underestimated as only vehicle-related crashes are reported. For now, very limited information about cyclists' distraction on fatality is available. Not only the limited data sources but also the number of studies that have been published on distracted biking is small. As a result, our knowledge about the effect of distraction on cycling is insufficient.

\subsection{Cyclist Cognitive Distraction}
Cyclist distraction refers to the phenomenon where a cyclist's focus and attention are diverted from the road and the cycling task. Previous studies have reported that distractions have a major prevalence among bike users and that they play a significant role in the prediction of the traffic crash rates of cyclists, through the mediation of risky behaviors \cite{useche2018distraction}. Studying cyclists' behaviors under the influence of distraction can provide evidence for interventions to address safety-related issues. 

Similar to diving distraction, cycling distraction can be categorized into three main types: visual (taking the eyes off the road), manual (taking the hands off the handlebar), and cognitive (taking the mind off cycling) \cite{stutts2001role}. Apart from cyclists' physical or psychological state changes like fatigue or stress, involvement in any type of secondary tasks (tasks that are not related to the cycling main task) is the main cause of distraction. This study will focus on cognitive distraction as it is related to the most frequently reported secondary task during cycling, such as listening to music or talking on phone through the earphones \cite{mwakalonge2014distracted,wolfe2016distracted}. This type of distraction can significantly impair a cyclist's ability to process information, respond to changing road conditions, and make safe decisions and maneuvers while cycling. For example, a study in real traffic examined the glance behavior of teenage cyclists while listening to music. The results indicated that a substantial percentage of participants cycling with music decreased their visual performance \cite{stelling2018study}. Another study found that listening to music can significantly increase cyclists' missing rate of auditory stop signals. Completing a task on the mobile phone (both handheld and hands-free) resulted in increased response time to an auditory stop signal and also reduced overall auditory perception \cite{de2011effects}. 

\subsection{Existing Methods for Studying Cognitive Distraction}
The current state of knowledge on cyclist distraction is mostly retrieved from surveys or observational studies. For example, an observational study in New York City shows that headphone use is the most prevalent distraction among local cyclists \cite{ethan2016analysis}. However, observational studies are unable to track cyclists' physiological changes, which helps to understand the amount of cognitive load. Additionally, observational studies do not provide the details of secondary tasks (e.g., headphone use can be either music listening or talking on the phone). Surveys from different areas around the world have been collected to study cyclists' distracted behavior, listening to music or talking with earphones have been identified as the most prevalent distractions \cite{terzano2013bicycling,wolfe2016distracted,young2020australian}. However, the subjective response collected in the surveys does not always reflect the participant's real-world response due to hypothetical bias \cite{fitch2018relationship}. Another way to collect data is the naturalistic study which records cyclists' responses in real-world conditions \cite{rupi2019visual}. Naturalistic study guarantees real-world data but it is limited by the potential safety risks, high costs, noise-diluted data, and difficulties in environmental control \cite{stelling2018study,fitch2020psychological,teixeira2020does}. To overcome the shortcomings of the existing methods, high-fidelity bike simulators have been developed by different researchers. The prevalence of virtual reality (VR) or Immersive Virtual Environments (IVE) technology in the past decade further provides a low-cost and controllable solution for experimental study to evaluate the responses of cyclists to different roadway designs and conditions \cite{guo2022orclsim}. 

The most frequent secondary tasks, both listening to music and talking with earphones can be categorized as cognitive distractions. One of the main challenges in the quantitative analysis of cognitive distraction is the difficulty in measuring the workload needed for certain tasks. To understand the mechanism of distraction, a standardized secondary task with different levels of workload is required in the experimental study. To our knowledge, no prior studies have applied such methods for cyclist distraction. In other research fields, several standardized secondary tasks have been developed to simulate different levels of workload. For instance, to simulate the phone conversation, an alternative mock
cellphone task was used in a driving-related study as a cognitive distraction \cite{ebadi2020impact}. The mock cellphone task was designed to simulate cognitive load when talking on the phone, and the impact of this type of task was reported to be similar to a hands-free cellphone conversation in a prior study \cite{muttart2007driving}.

\subsection{Physiological Responses under Cognitive Distraction}
To objectively measure cognitive distraction, physiological responses can be useful metrics to record and monitor within an experiment. Previous studies have used physiological responses as a proxy to detect distraction in humans \cite{lohani2019review}. Specifically, prior studies in transportation engineering, mostly in driving conditions, have used eye gaze patterns \cite{azman2010non,hua2021effect,musabini2020heatmap,he2020time}, EEG \cite{strayer2015assessing,sena2016experimental}, electrodermal activity (EDA) \cite{solovey2014classifying,he2020time,steinberger2017road}, heart rate (HR) \cite{tavakoli2022multimodal,tavakoli2022driver,solovey2014classifying,azhar2022development,steinberger2017road}, heart rate variability (HRV) \cite{solovey2014classifying,steinberger2017road}, and skin temperature to detect drivers' distraction\cite{yusoff2017selection}. Here, we direct our focus to cardiovascular and skin-related metrics for this research. Cardiovascular metrics, including HR and HRV, can be retrieved through multiple devices such as an Electrocardiogram (ECG) and  Photoplethysmography (PPG), where PPG is the technology that is mostly used in smart wearable devices and can be applied in both naturalistic and experimental conditions \cite{tavakoli2021harmony}. Based on the heart signal retrieved through either of the aforementioned devices, features of heart activity can be calculated that are generally referred to as heart rate variability (HRV) features. These features are results of signal processing applied to the heart signal in time, frequency, and nonlinear domain. An example of the time domain HRV is the root mean squared of the successive intervals of heart beats (R-R intervals) referred to as RMSSD. An example of the frequency domain HRV metric is the power of the high frequency (between 0.15 to 0.4 Hz) and low frequency band (between 0.04 to 0.15 Hz) of the HRV spectrum referred to as HF and LF. Lastly, an example of the nonlinear domain HRV feature is the entropy of the beat-to-beat intervals. Previous studies have showed that these features are correlated with certain human states such as stress level (correlated with decrease in RMSSD) and cognitive load ( increase in HF) \cite{charles2019measuring,solhjoo2019heart,lohani2019review,kim2018stress}.   

Electrodermal activity (EDA) or sometimes referred to as Galvanic Skin Response (GSR) have also been shown to be correlated with certain human states such as stress and workload. EDA signals are generally first decomposed into two main components of tonic and phasic. Tonic is the long-term changes in the signal, whereas phasic is the momentary changes in the EDA signal. The tonic level is then used to define the skin conductance level (SCL), and the phasic component is used to define the skin conductance responses (SCR). Both SCL and SCR have been shown to be correlated with higher cognitive load and stress level \cite{lohani2019review}. 

Within bicycle research, application of physiological sensors has mostly been centered around detecting stress, comfort, and emotions \cite{guo2021benchmarking,guo2023psycho}. Additionally, some naturalistic studies have been conducted to evaluate bicyclists' behavior and physiological responses in different contextual settings (\cite{guo2021benchmarking,rupi2019visual,teixeira2020does}). These preliminary studies revealed that psycho-physiological metrics (e.g., heart rate (HR), gaze variability, and skin conductance) are indicators of how cyclists' perceive change in different contextual settings. While previous studies have provided significant insights into using physiological responses for distraction detection in driving as well as cyclists state (e.g., stress and comfort) detection, they have not been used for cyclists cognitive distraction defections much. In other words, we still have very a limited understanding of bicyclists' physiological responses in different levels of cognitive distraction, especially in IVE studies.

\subsection{Research Goals}
The goal of this experiment is to study the effect of cognitive distraction on cyclist behavior. More specifically, we are interested in applying the standardized secondary task in the IVE bicycle simulator to simulate different levels of cognitive workload, and explore cyclists' physiological responses in different situations. 

The research hypothesis are 
\begin{enumerate}
    \item Listening to high-tempo music and talking on the phone can be candidates for standardized secondary tasks and can be used to simulate different cognitive distractions in the IVE. 
    \item Listening to high-tempo music and talking on the phone result in varying levels of behavioral performance such as lower speed in comparison to a baseline of no distraction-biking.  
    \item Listening to high-tempo music and talking on the phone result in significantly different physiological responses such as different gaze/head movement variation, HRV, and EDA in comparison to baseline no-distraction condition.
\end{enumerate}

\section{Methodology}
The study is reviewed and approved by the Institutional Review Board for the Social and Behavioral Sciences from University of Virginia (IRB-SBS Protocol \# 2148). All experiments were performed in accordance with relevant named guidelines and regulations. Informed consent was obtained from all participants and/or their legal guardians.

\subsection{Experiment Design} 
This research studies the effect of cognitive distraction on cyclist behavior in the proposed IVE bicycle simulator framework. The cognitive distraction will be triggered by both the standardized secondary task (mock phone conversation task) and the actual task (music listening). Each participant in this within-subject design will experience 3 different conditions (baseline, music listening, and mock phone conversation) in random order. 

\subsubsection{Cognitive Secondary Tasks - Mock Phone Conversation Task}
As an alternative to a real cellphone conversations, a mock cellphone conversation task was used in this study as a cognitive distraction, particularly because typical conversations are difficult to experimentally control. While performing the distraction task, participants were instructed to listen to a series of generic English language sentences synonymous with the previously validated grammatical reasoning task and respond aloud to the subject, object, and whether the sentence was plausible or not. The experimenter would remotely initiate the task with a button press at the beginning of each scenario and similarly, terminate it a few seconds before the end of every scenario. The participants listening to each sentence were then asked to reply aloud: the subject of the sentence, the object of the sentence, and whether or not the sentence was plausible. For example, for the sentence, \textit{“A child jumped a rope,”} the correct answer is \textit{“Child, Rope, and Yes.”} Similarly, an implausible sentence would be, “A cat baked the television,” and the correct answer would be: “Cat, Television, and No.” The mock cellphone task was designed in such a way that the cyclist would be cognitively loaded, and as found in prior studies, the impact of this type of task would be similar to that of a hands-free cellphone conversation \cite{ebadi2020impact}.

\subsubsection{Cognitive Secondary Tasks - Music Listening Task}
In addition to the standardized secondary tasks, an actual secondary task, music listening is considered in this experiment as well. In the music listening condition, the participants will be asked to listen to a popular song of the year 2022, which is a positive song with a tempo of 174 Beat Per Minute. The track lasts for 3:05. The song will be played during cycling automatically by the experimenter, the participants can hear the song from the earphone of the VR headset, similar to listening to music with earphones in the real world.

\subsection{Road Environment in IVE}
\begin{figure} 
    \centering
    \includegraphics[width=\linewidth]{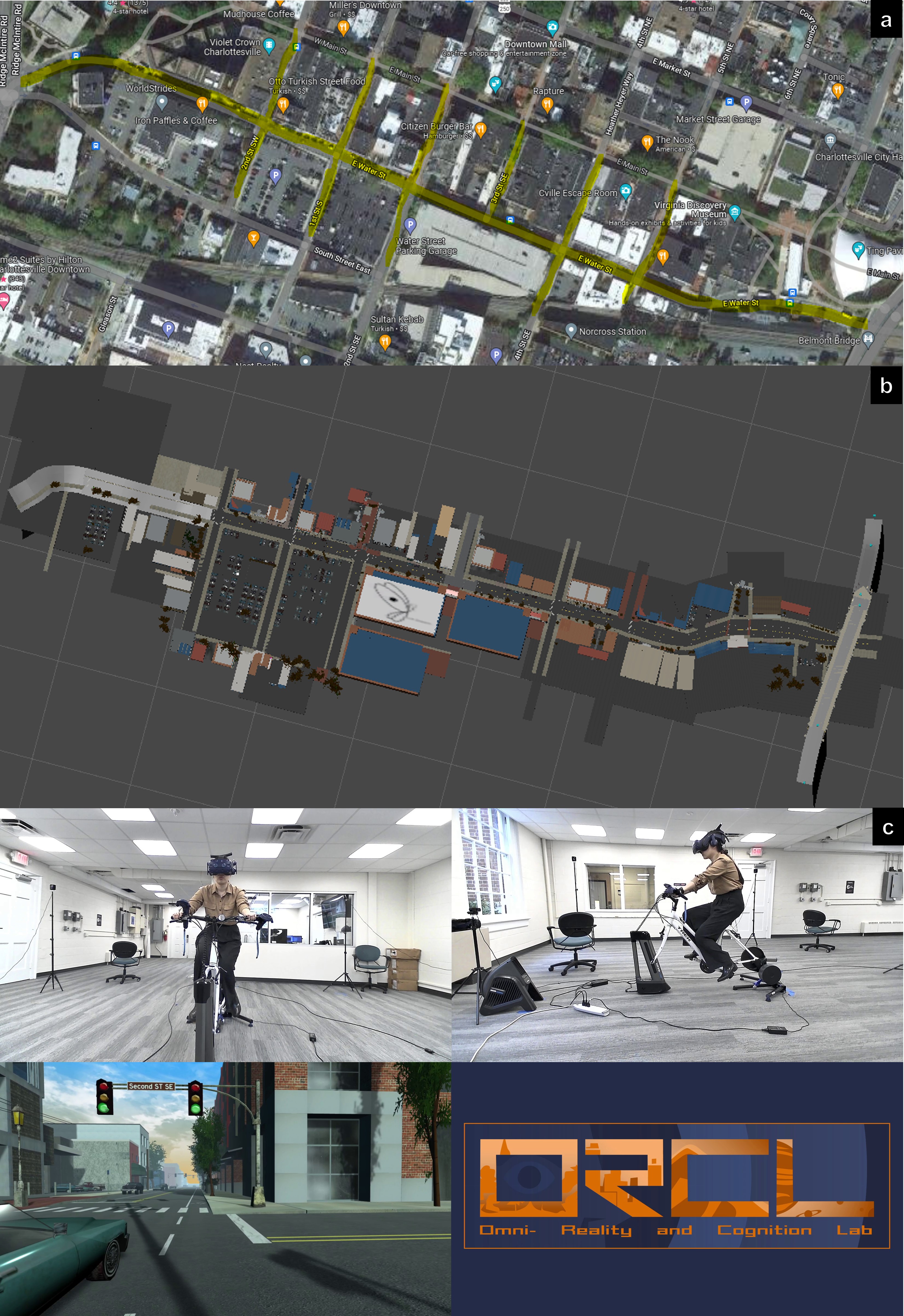}
    \caption{Illustration of the IVE road environment and bike simulator. (a) Bird view from the map of the real road. The built road segments are highlighted in yellow. (b) Bird view of the created water street area within the IVE in Unity software. (c) Snapshot of video collection system with room cameras and VR headset recording}
    \label{fig:IVE_road_environment_cognitive}
\end{figure}

The IVE is developed on a 1:1 scale in Unity 3D game engine and steamVR platform, based on the Water Street corridor in Charlottesville, Virginia. Water Street is well-trafficked by bicyclists and is being considered for redesign by the city of Charlottesville due to the crash history in this area as shown in Fig.\ref{fig:IVE_road_environment_cognitive}(a). The IVE road starts from the intersection of West Main Street and Ridge Street and ends at the intersection of East Water Street and 9th Street NE (the Belmont Bridge). Bike lanes are designed for the road in the IVE with a standard bike lane width of 4 feet (1.2m).

\subsection{Apparatus}
In addition to the virtual environment simulation in Unity, several software or devices have been utilized in this study as listed below, which is based on our prior framework introduced in Guo et al 2022 \cite{guo2022orclsim}:

\subsubsection{VR devices}
The HTC VIVE Pro Eye headset is connected wirelessly to the control computer in the lab. Two controllers are attached to the handlebar of the simulator, and the spatial location of the controllers is detected to reflect the turning movements. The squeezing value of the trigger button on the back side of the right controller is programmed for the braking of the bike. 
With the C\# scripts written in the Unity scenario, the cycling performance data such as speed, lane position, head movement, and braking data are collected from the headset and controllers. 

\subsubsection{Bike simulator}
As can be viewed from Fig.\ref{fig:IVE_road_environment_cognitive}(c), an average physical Trek Verve bike (without wheels) is assembled with a series of Wahoo Kickr Smart Trainer (Climb, Headwind, and ANT+) to receive pedaling power, and simulate resistance/headwind feedback based on current cycling speed. With the bike trainer, the pedaling input power data is available after each play.

\subsubsection{Smartwatch}
The Empatica E4 smartwatch is used to collect participants' heart rate data and EDA data. The smartwatch is connected to a smartphone via Bluetooth. The data collection is initialized and controlled by the smartphone. The clock on the smartphone is synchronized with the control computer before each participant's test. 

\subsubsection{video collection}
As shown in Fig.\ref{fig:IVE_road_environment_cognitive}(c), the OBS studio software integrates the three video collection components: two video recordings from cameras capturing the body position of the participant and one screen recording of the participant’s point of view in IVE.

\subsection{Experiment Procedure} 
After signing the consent form, each participant was asked to put on a smartwatch before completing the pre-experiment survey. After finishing the pre-experiment survey, instructions are given on how to use the VR headset, controllers, and bike simulator. After the bike is adjusted to a comfortable position, the participant mounts the bike and puts on the headset. Next, the participant is guided through the eye tracker calibration. After the IVE system setup, the participant is placed into a familiarization scenario (without any vehicle traffic) to become accustomed to interacting with the IVE. In this environment, the participant can practice pedaling, steering, and braking and the practice procedure can be repeated until the participant feels comfortable. If the participant feels any motion sickness, they may stop the experiment at any point and still receive compensation for participation. 

Once the participant is comfortable in the training environment, they experience the three design scenarios in random order, where each experiment trial lasts about two minutes, with a two-minute break between each scenario. Once the participant has completed all three scenarios, they are asked to complete the post-experiment survey. On average, each participant spends 30 minutes completing the experimental procedure.

\subsection{Participants}

75 participants were recruited for the experiment. Among them 40 are female, 33 are male,  1 participant identified as other and 1 participant didn't provide gender information. Most of the participants are local bicyclists, students, and faculty members from the University of Virginia. All participants are 18 or older and without color blindness. The mean age is 24.5 with a standard deviation of 4.7, and the median age is 24.5 years old as well (one participant didn't provide the age information). 



\subsection{Physiological Data Analysis}
In order to analyze the physiological data, we have taken advantage of multiple modeling techniques and packages scripted in python. 


\subsubsection{Heart Rate Variability (HRV)}
Based on the interbeat interval data (IBI) recorded through Empatica E4, we can calculate HRV features. As mentioned previously, these features span across different domain of time, frequency, and nonlinear. We use the pyHRV \cite{gomes2019pyhrv} package scripted in python to calculate these features in all three domains for each participant during the experiment (e.g., HF, and RMSSD). Note that some of the calculated feature may not be applicable to short term data collection (e.g., LF), thus we only focus on the features that are suitable for a short period of time.

\subsubsection{Electrodermal Activity (EDA)}
In order to calculate the tonic and phasic compositions of the EDA signals, we first denoise the EDA signal recorded through Empatica E4 by passing it through a Butterworth signal with a high pass of 10 Hz. We then feed the resulting signal to the well-known Neurokit 2 package \cite{Makowski2021neurokit}. This package allows for decomposing the signal to tonic and phasic component and calculating a variety of EDA related features such as number of phasic skin conductance responses and skin conductanc elevel (SCL). We compare number of SCR peaks across different participants and conditions in the study.

\subsection{Statistical Analysis}
In order to compare across conditions, we use Linear Mixed Effect Models (LMM), for their capability in addressing individual differences. Different people have various baseline values for their physiological signals. For instance, a person's baseline HR might be at 60 bpm, while another person's might be at 80 bpm. Additionally, the slope of change in a peconductancealevel signals across different conditions can be very different relative to another participant. LMM is similar to linear regression in measuring the main effects in a study but with a difference that it accounts for random changes across participants referred to as random factors \cite{fox2002linear,brown2021introduction} through random intercept and random slope. An LMM is defined as follows:

\begin{equation}\label{lmm}
    y = X\beta + bz +\epsilon
\end{equation}

In equation (\ref{lmm}), $y$ is the dependent variable (e.g. number of skin conductance responses), $X$ is the matrix of predictor variables (type of distraction), $\beta$ is defined as the vector of fixed-effect regression coefficients, $b$ is defined as the matrix of random effects, $z$ is defined as the coefficients for each random effect, and $\epsilon$ is the error term. Additionally, we can define the elements of the $b$ and $\epsilon$ matrices as follows:
\begin{equation}
    b_{ij} \sim N(0,\psi_k^{2}),Cov(b_k,b_{k'})
\end{equation}

\begin{equation}
    \epsilon_{ij} \sim N(0,\sigma^{2}\lambda_{ijj}),Cov(\epsilon_{ij},\epsilon_{ij'})
\end{equation}

We use lme4 package in R \cite{bates2007lme4} for applying LMM to the data. 

\section{Results}
This section reports the results of the experiment. The following subsections describe the summary statistics (from the pre-and post-experiment surveys), the bicyclists’ physical behavior (cycling speed, input power, and lateral lane position), and physiological responses in different roadway designs.

\subsection{Survey Response}
\subsubsection{Pre-experiment Survey}
Participants' attitude towards cycling is collected in the pre-experiment survey. 74 participants answered the question about what types of bicyclists they are. The majority of the participants have a positive attitude towards cycling, 6 participants (8.11\%) indicated their attitude towards cycling as "No way, no how" - I do not ride a bike, 25 participants (37.84\%) identified themselves as "Interested but Concerned" - I like the idea of riding but have concerns. The rest of the participants had a higher preference for cycling as 28 (33.78\%) indicated themselves as "Enthused and Confident" -I like to ride and will do so with dedicated infrastructure and the remaining 15 (20.27\%) chose "Strong and Fearless" - I will ride anywhere, no matter the facilities provided.

\begin{figure} 
    \centering
    \includegraphics[width=\linewidth]{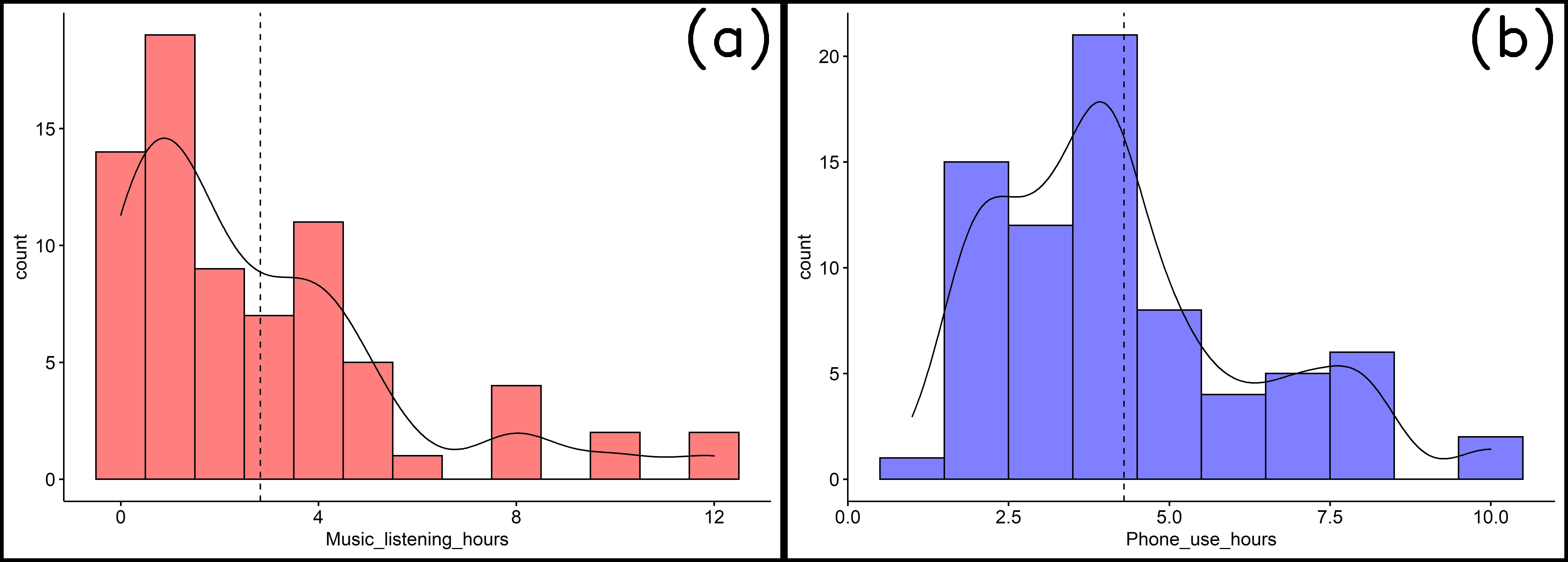}
    \caption{Participants' daily hours spend in music listening (a) and phone usage (b)}
    \label{fig_cognitive:Secondary_task_hours}
\end{figure}

The engagement of secondary tasks both in daily life and during cycling is also collected in the pre-experiment survey. We asked the participants to estimate how many hours they spend in music listening and phone usage, as well as the frequency of music listening, and phone talking when they ride a bike. The distribution of daily hours spend on music listening and phone are displayed in Figure \ref{fig_cognitive:Secondary_task_hours}. The average music listening hours are 2.82 hours (sd = 2.88 hours). And unsurprisingly, the average hours spent on the phone is higher (mean = 4.30 hours, sd = 2.07 hours). Participants' music listening and phone talking frequency while riding a bike is self-reported with 5-point Likert scales, with 5 options of "Never ($<$10\%)", "Seldom (about 25\%)", "Sometimes (about 50\%)", "Often (about 75\%)", and "Always ($>$90\%)". Participants were asked to choose an option that is closest to them. Table \ref{table:music_phone_frequency} summarizes the results for these two questions. The participants have a higher frequency of music listening than talking on the phone while biking. More than half of the participants admitted that they have music-listening behavior while biking, and only about 25\% of the participants reported that they had the experience talking on the phone while biking before.

\begin{table}[h!]
\caption{Summary of participants' music listening and phone talking frequency while riding a bike}
\label{table:music_phone_frequency}
\centering
 \begin{tabular}{c c c c } 
  \hline
 Secondary Task Type & Frequency  & Number  & Percentage \\ 
 \hline
 & Never ($<$ 10\%) & 36 & 48 \\ 
            & Seldom (about 25\%) & 9 & 12\\ 
    Music Listening      & Sometimes (about 50\%) & 7 & 9.3 \\
   while Biking    & Often (about 75\%) & 15 & 20 \\
      & Always ($>$ 90\%) & 7 &  9.3 \\
            & NA & 1 & 1.3\\
  \hline
 & Never ($<$ 10\%) & 55 & 73.3 \\ 
            & Seldom (about 25\%) & 18 & 24\\ 
    Talking on the Phone      & Sometimes (about 50\%) & 1 & 1.3 \\
   while Biking    & Often (about 75\%) & 0 & 0 \\
      & Always ($>$90\%) & 0 &  0 \\
            & NA & 1 & 1.3\\
 \hline
 \end{tabular}
\end{table}

\subsubsection{Post-experiment Survey}
In the post-experiment survey, participants' stated preferences over the three scenarios are collected in three aspects: safety, comfort and distraction. For each question, the answer is to choose their subjective ratings from a 5-point Likert scale. 

\begin{figure} 
    \centering
    \includegraphics[width=0.8\linewidth]{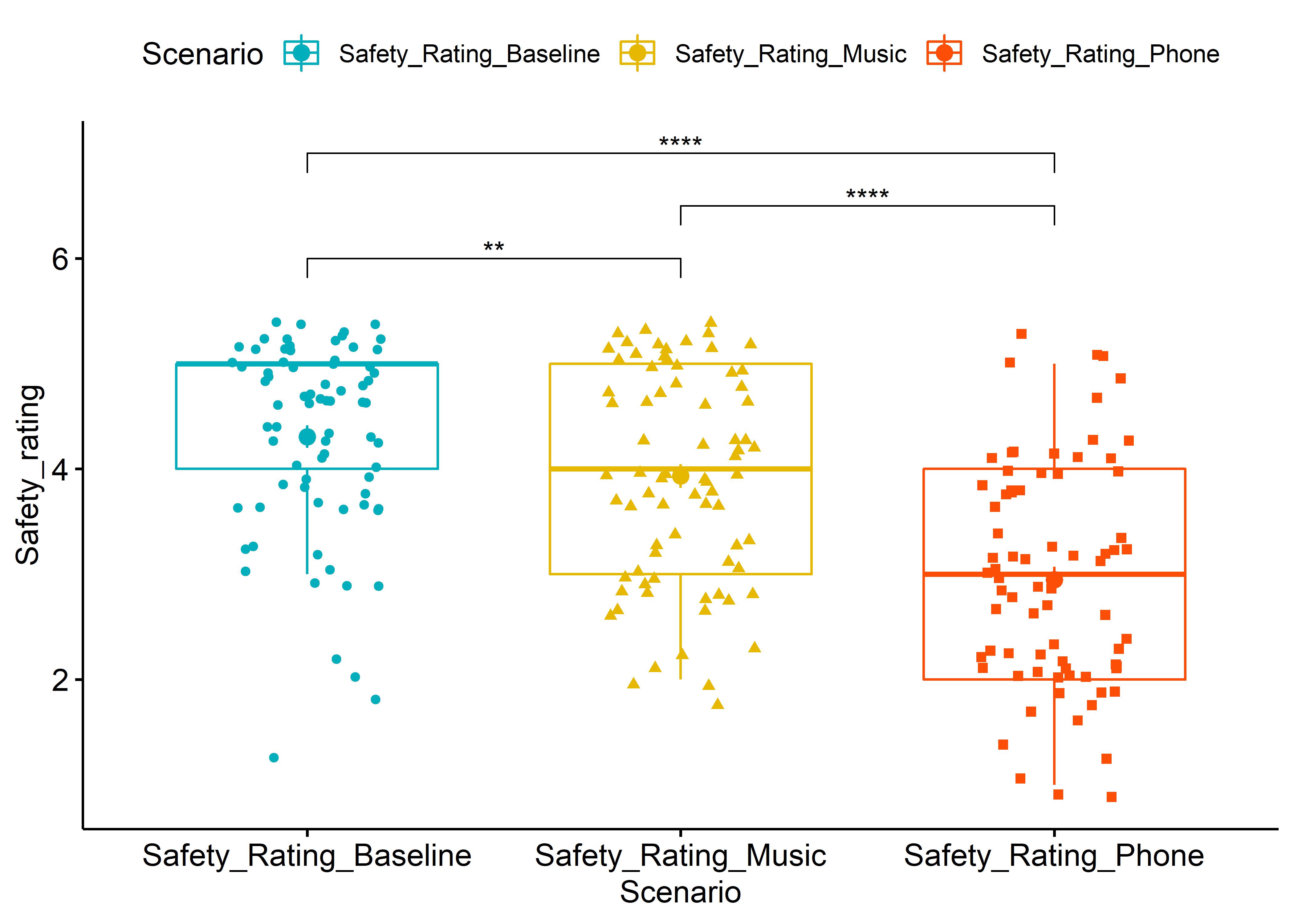}
    \caption{Subjective safety rating of different scenarios (jitter points to avoid overplotting)}
    \label{fig_cognitive:safety_rating}
\end{figure}

For subjective safety rating, the Baseline scenario is rated as the safest scenario with an average score of 4.31/5.0, followed by the Music Listening (3.93/5.0), then the Mock phone conversation scenario (2.95/5.0), the differences between all the three scenarios are significant, as shown in Figure \ref{fig_cognitive:safety_rating} (Baseline v.s. Music listening, p = 0.00297; Baseline v.s. Mock phone conversation, p $<$ 0.0001; Music listening v.s. Mock phone conversation, p $<$ 0.0001). 

\begin{figure} 
    \centering
    \includegraphics[width=0.8\linewidth]{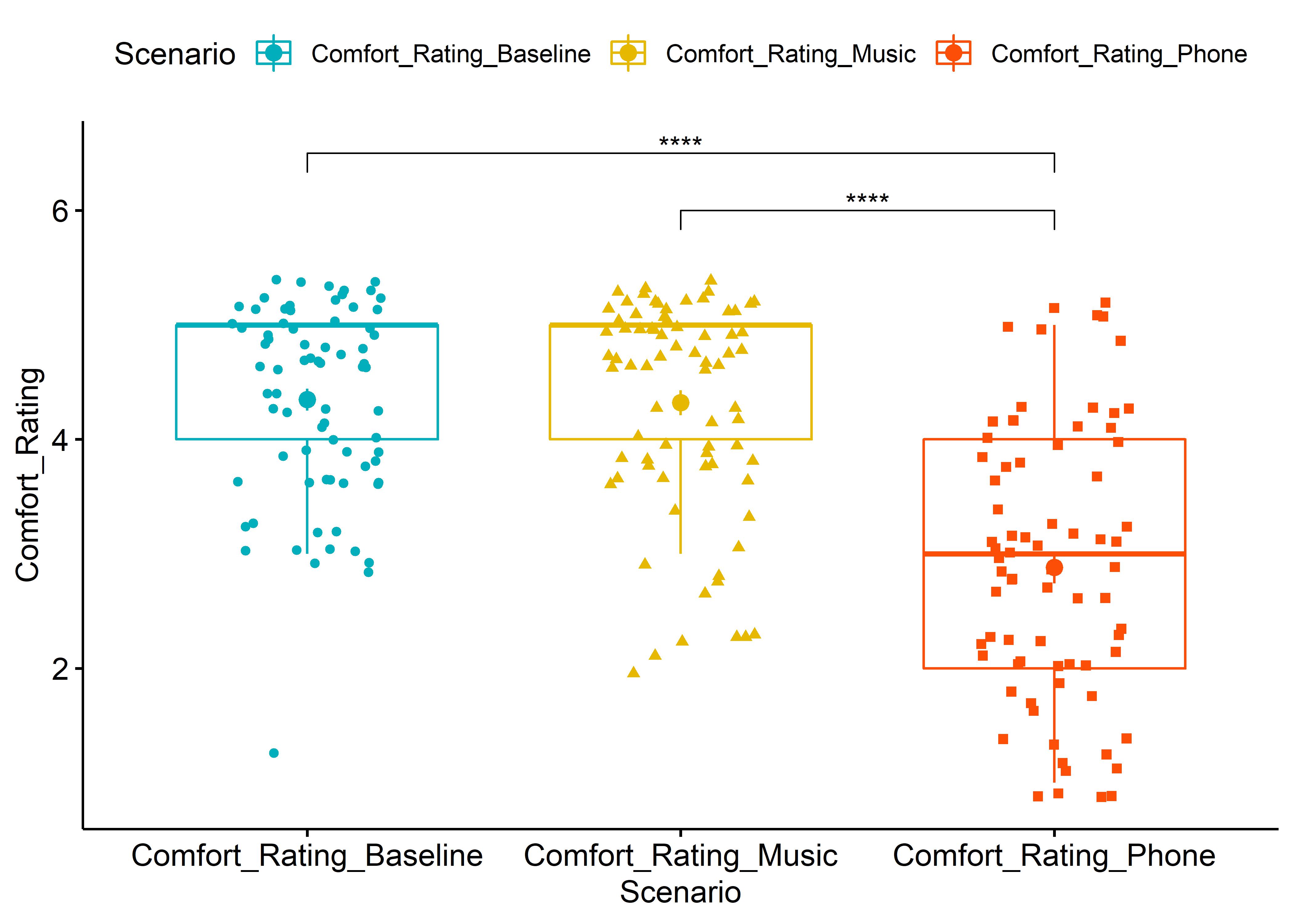}
    \caption{Subjective comfort rating of different scenarios (jitter points to avoid overplotting)}
    \label{fig_cognitive:comfort_rating}
\end{figure}

For subjective comfort rating, the scores of the Baseline (4.35/5.0) and Music listening (4.32/5.0) scenarios are close to each other, and both are significantly higher than the Mock phone conversation scenario (2.88/5.0), with both p values smaller than 0.0001, as shown in Figure \ref{fig_cognitive:comfort_rating}.

\begin{figure} 
    \centering
    \includegraphics[width=0.8\linewidth]{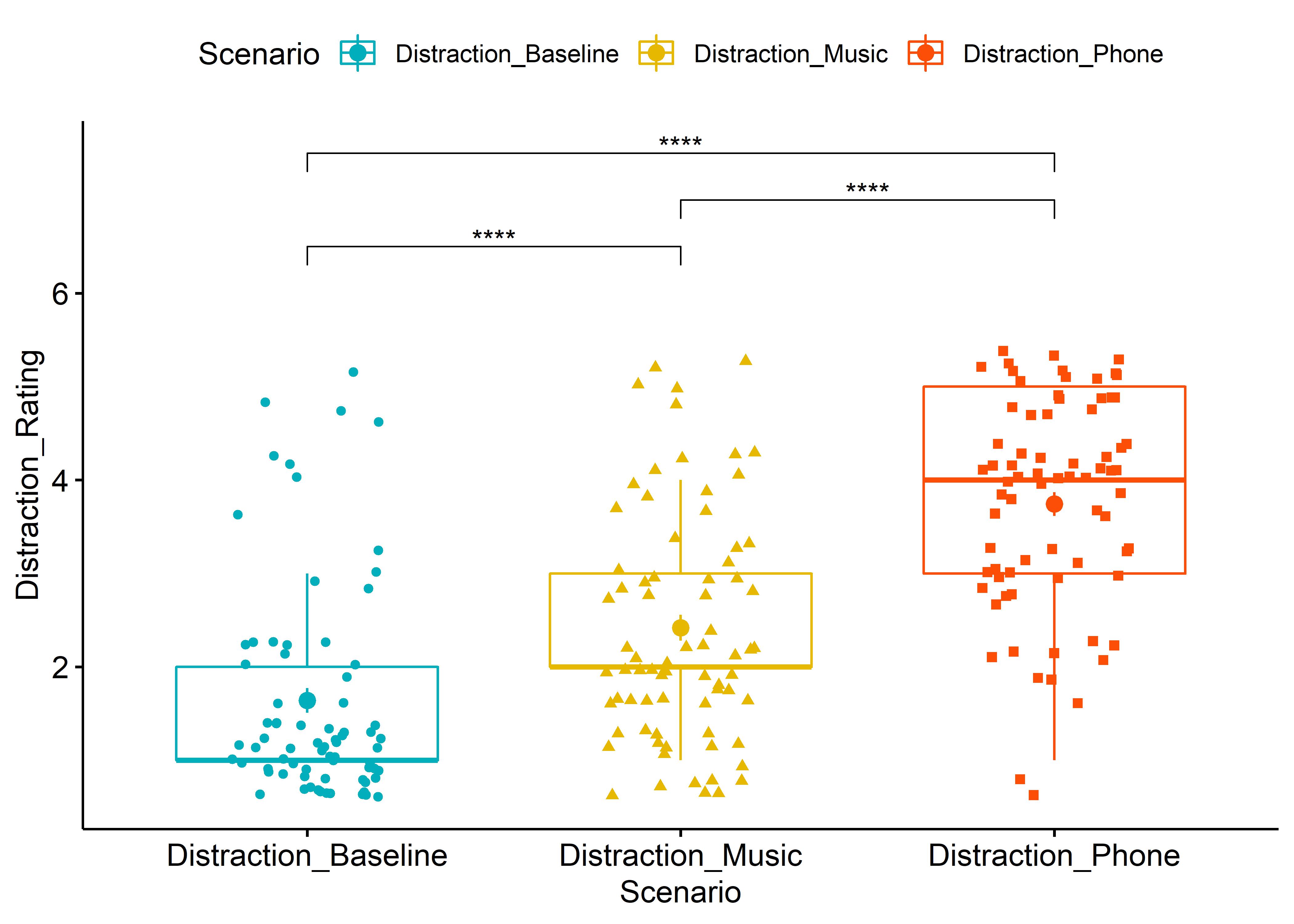}
    \caption{Subjective distraction rating of different scenarios (jitter points to avoid overplotting)}
    \label{fig_cognitive:distraction_rating}
\end{figure}

For subjective distraction rating, the Mock phone conversation scenario is rated as the most distracting scenario with an average score of 3.74/5.0, followed by the Music Listening (2.42/5.0), then the Mock phone conversation scenario (1.64/5.0), the differences between all the three scenarios are significant, as shown in Figure \ref{fig_cognitive:distraction_rating} (all the p values are smaller than 0.0001). 

\subsection{Cycling Performance}
\subsubsection{Speed}
\begin{figure} 
    \centering
    \includegraphics[width=\linewidth]{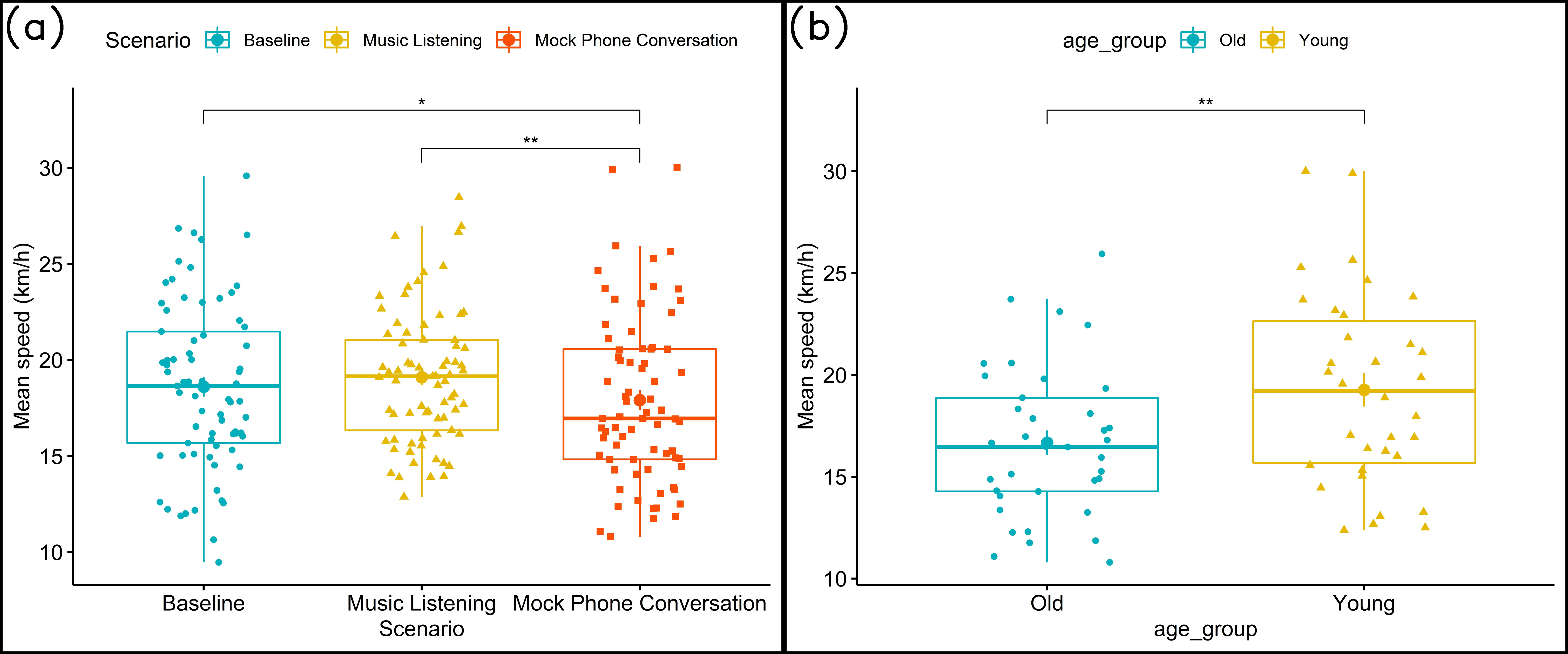}
    \caption{Mean speed of (a) different scenarios, (b) different age groups in the Mock phone conversation scenario}
    \label{fig_cognitive:mean_speed}
\end{figure}

For the mean speed, as indicated in Figure \ref{fig_cognitive:mean_speed}-a, there is a significant difference between the Baseline and the Mock phone conversation scenarios ($\beta = -1.262, SE = 0.435, p = 0.0428$) and between the Music listening and the Mock phone conversation scenarios ($\beta = 1.178, SE = 0.312, p = 0.0007$). Bicyclists' mean speed in the three scenarios (Baseline, Music listening, and Mock phone conversation) are 18.6 km/h, 19.1 km/h, and 17.9km/h, respectively. Age group differences are found only in the Mock phone conversation scenario, where the younger group (19.3 km/h) has a significantly higher cycling speed ($\beta = 2.903, SE = 1.058, p = 0.00783$) than the older group (16.7 km/h), as indicated in Figure \ref{fig_cognitive:mean_speed}-b. 

\begin{figure} 
    \centering
    \includegraphics[width=\linewidth]{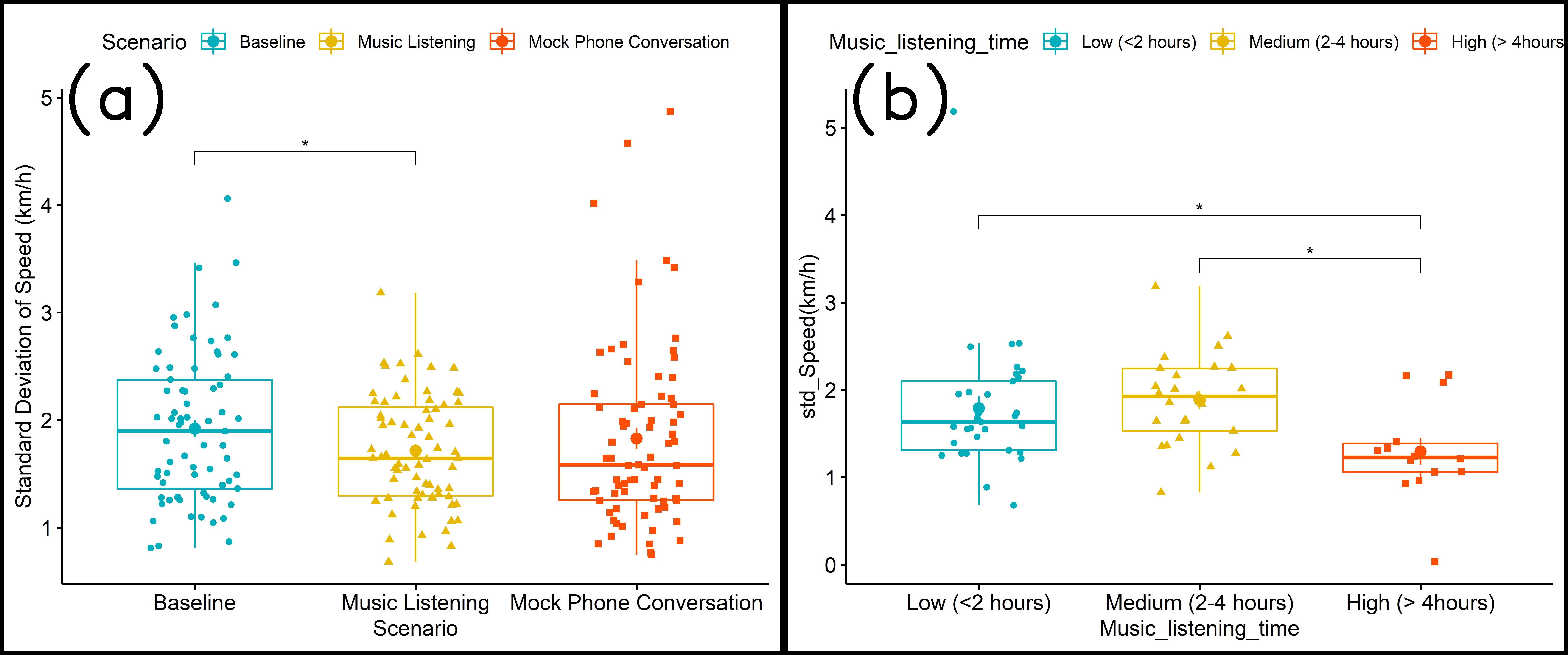}
    \caption{Standard deviation of the speed of (a) different scenarios, (b) participants with different music listening time in the Music listening scenario}
    \label{fig_cognitive:std_speed}
\end{figure}

For the standard deviation of speed, the results show there is a significant difference between the Baseline and the Music listening scenarios ($\beta = -0.30669, SE = 0.14391, p = 0.0348$), as shown in Figure \ref{fig_cognitive:std_speed}-a. Bicyclists' standard deviation of speed in the three scenarios (Baseline, Music listening, and Mock phone conversation) are 1.92 km/h, 1.74 km/h, and 1.83 km/h, respectively. For the Music listening scenario, it is found that participants who listen to music a lot (>4 hours daily) have a lower standard deviation of speed	($\beta = -0.572, SE = 0.226, p = 0.0142$), as shown in Figure \ref{fig_cognitive:std_speed}-b.

\subsubsection{Lateral Lane Position}
\begin{figure} 
    \centering
    \includegraphics[width= \linewidth]{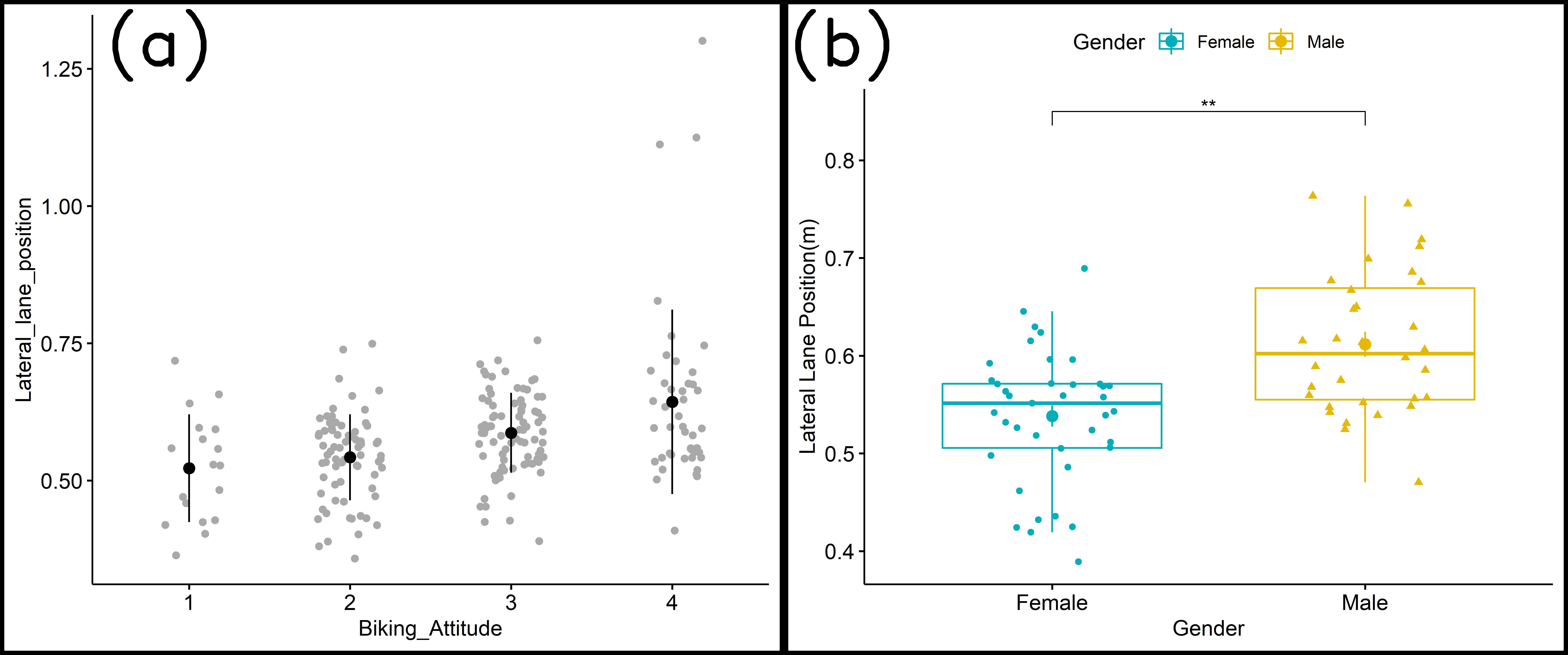}
    \caption{Mean lateral lane position of (a) different attitudes toward cycling, a higher score means a more positive attitude towards cycling. 1-4 indicates 'No way, no how', 'Interested but Concerned', 'Enthused and Confident', and 'Strong and Fearless', respectively. (b) different genders in the Music listening scenario}
    \label{fig_cognitive:mean_lane_position}
\end{figure}

For the lateral lane position, no significant differences are found between different scenarios. The average lateral lane position for the three scenarios is Baseline - 0.575m, Music listening - 0.569m, and Mock phone conversation - 0.592m. However, significant differences are found between the participants with different attitudes toward cycling. As can be seen from Figure \ref{fig_cognitive:mean_lane_position}-a, participants who hold more positive attitudes toward cycling will go more on the left (closer to the vehicle lane).

\subsubsection{Input Power}
\begin{figure} 
    \centering
    \includegraphics[width=\linewidth]{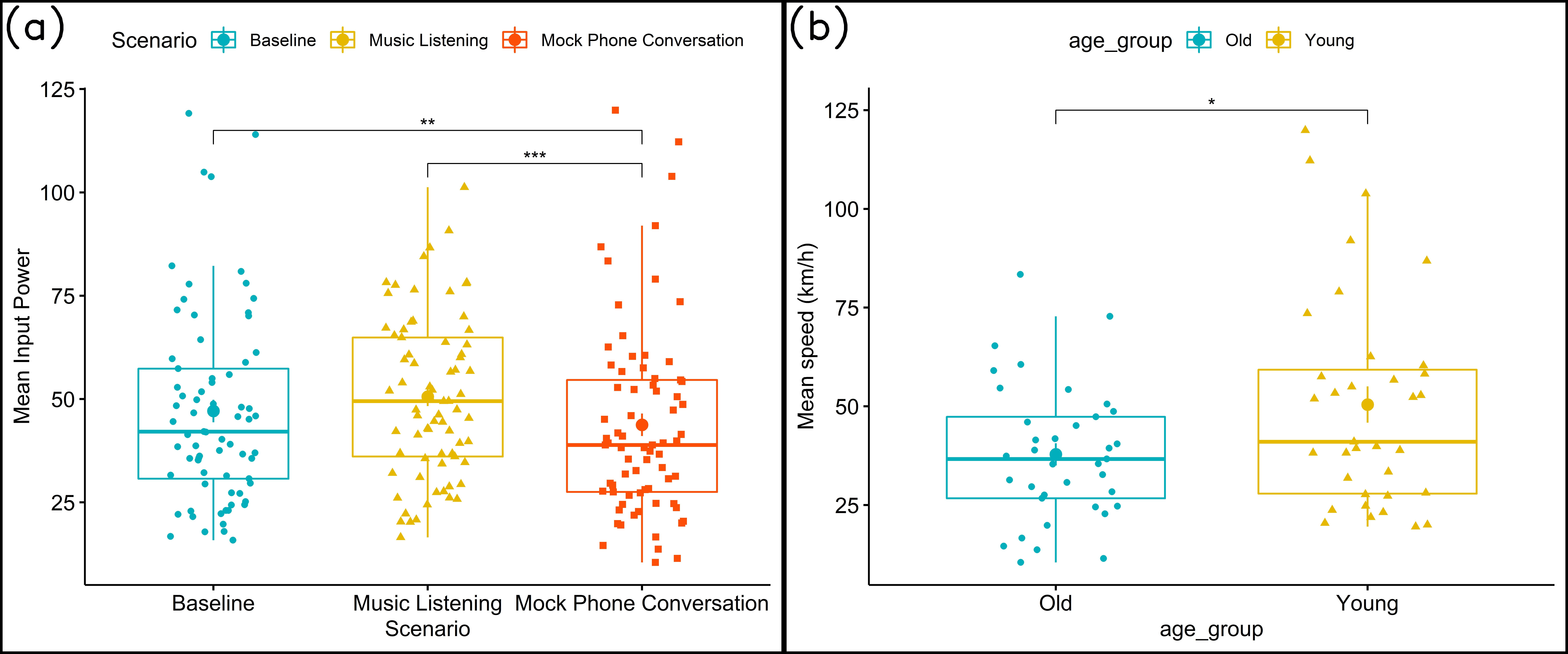}
    \caption{Mean input power of (a) different scenarios, (b) different age groups in the Mock phone conversation scenario}
    \label{fig_cognitive:mean_inputpower}
\end{figure}

For the mean input power, the average input power in music listening (50.6 Wattage) is 16\% higher than mock phone conversation (43.8 Wattage), and 7.5\% higher than baseline (47.1 Wattage). There is a significant difference between the Baseline and the Mock phone conversation scenarios ($\beta = -6.658, SE = 2.443, p = 0.00725$) and between the Music listening and the Mock phone conversation scenarios ($\beta = 6.72, SE = 1.75, p = 0.0006$), as shown in Figure \ref{fig_cognitive:mean_inputpower}-a. Age group differences are found only in the Mock phone conversation scenario, where the younger group (50.4 Wattage) has a significantly higher cycling input power ($\beta = 11.864, SE = 5.712, p = 0.0418$) than the older group (37.8 Wattage), as indicated in Figure \ref{fig_cognitive:mean_inputpower}-b.

\subsubsection{Head Movement}

\begin{figure} 
    \centering
    \includegraphics[width=\linewidth]{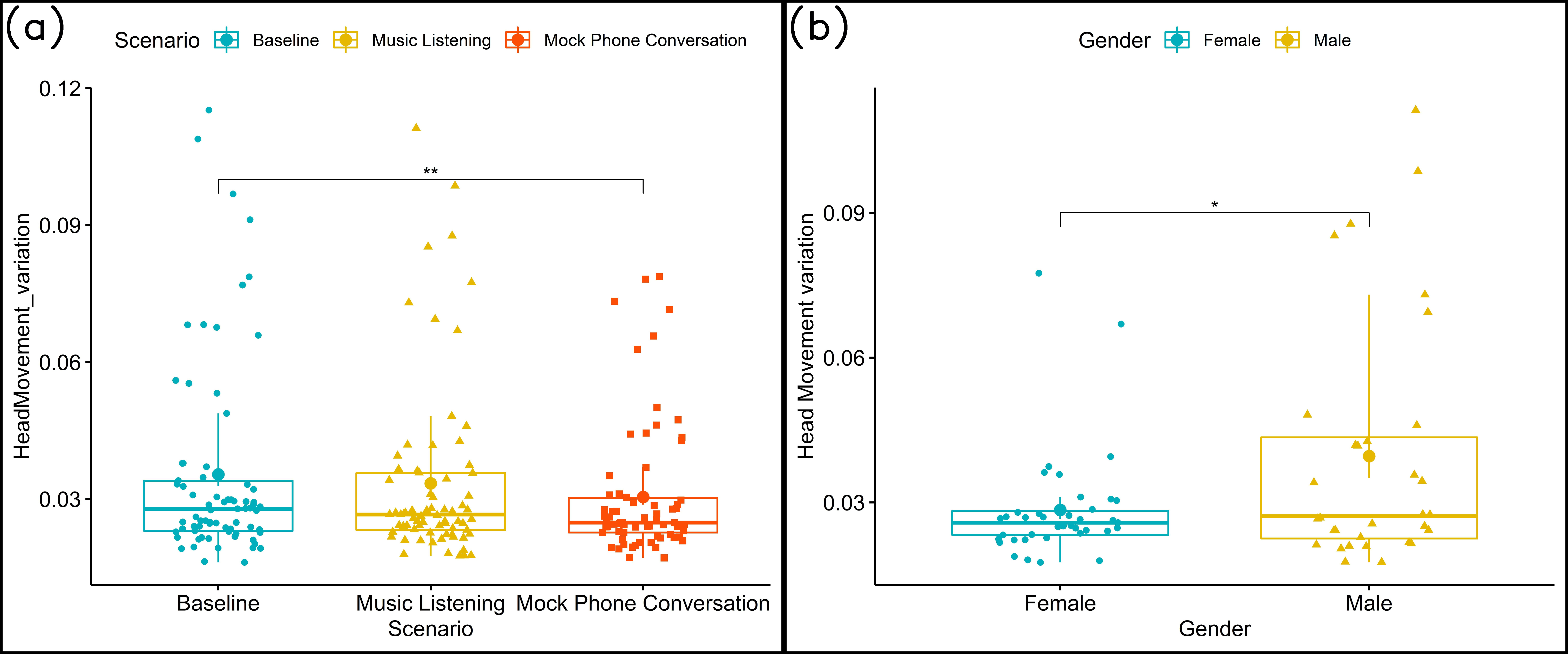}
    \caption{Head movement variation (a) different scenarios; (b) gender differences in music listening scenario}
    \label{fig_cognitive:HeadMovement}
\end{figure}

The head movement result shows a significant difference between the Baseline and the Mock phone conversation scenario with $\beta = -0.00636, SE = 0.00220, p = 0.00435$, as shown on Figure \ref{fig_cognitive:HeadMovement}-a, participants had a lower variation of head movement direction in the Mock phone conversation scenario than the Baseline. Additionally, in the music listening scenario, male participants have a higher head movement variation than female participants with $\beta = 0.0122, SE = 0.00538, p = 0.0266$ (Figure \ref{fig_cognitive:HeadMovement}-b).

\subsection{Physiological Response}
\subsubsection{HR/HRV}
The comparison of the mean heart rate across conditions indicates that there are no significant differences between the three scenarios with a 95\% confidence level. The mean HR (beat per minute) of the Baseline, Music listening, and Mock phone conversation are 92.89, 92.07, and 90.66 bpm, respectively.

\begin{figure} 
    \centering
    \includegraphics[width=0.8\linewidth]{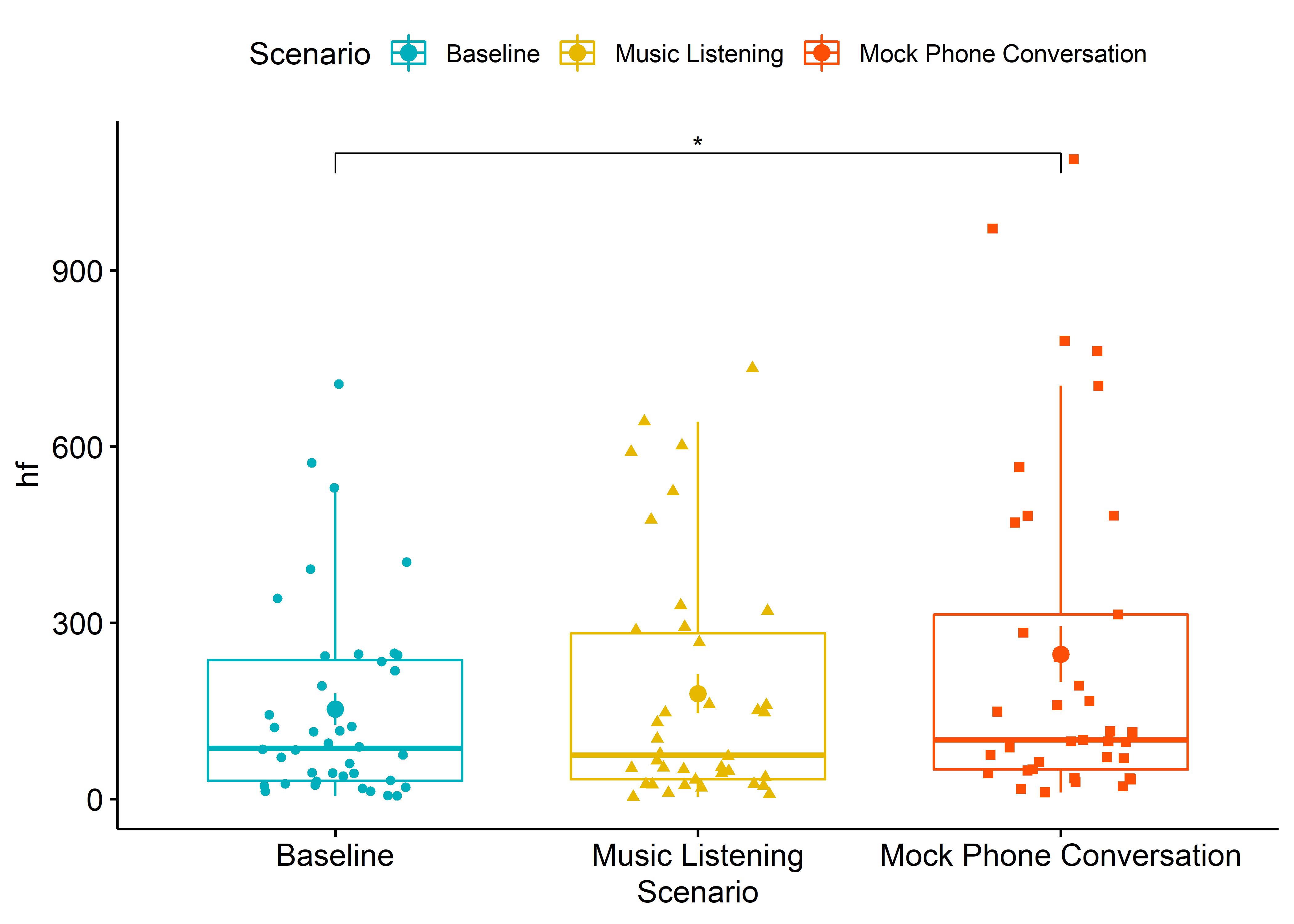}
    \caption{HF-HRV features in different scenarios}
    \label{fig_cognitive:HF-HRV}
\end{figure}

For HF-HRV, the LMM model results show that cyclists in the Mock phone conversation scenario have higher HF-HRV than the baseline ($\beta = 232.31, SE =  114.40, p = 0.0454$), as indicated in Figure \ref{fig_cognitive:HF-HRV}. 

\begin{figure} 
    \centering
    \includegraphics[width=\linewidth]{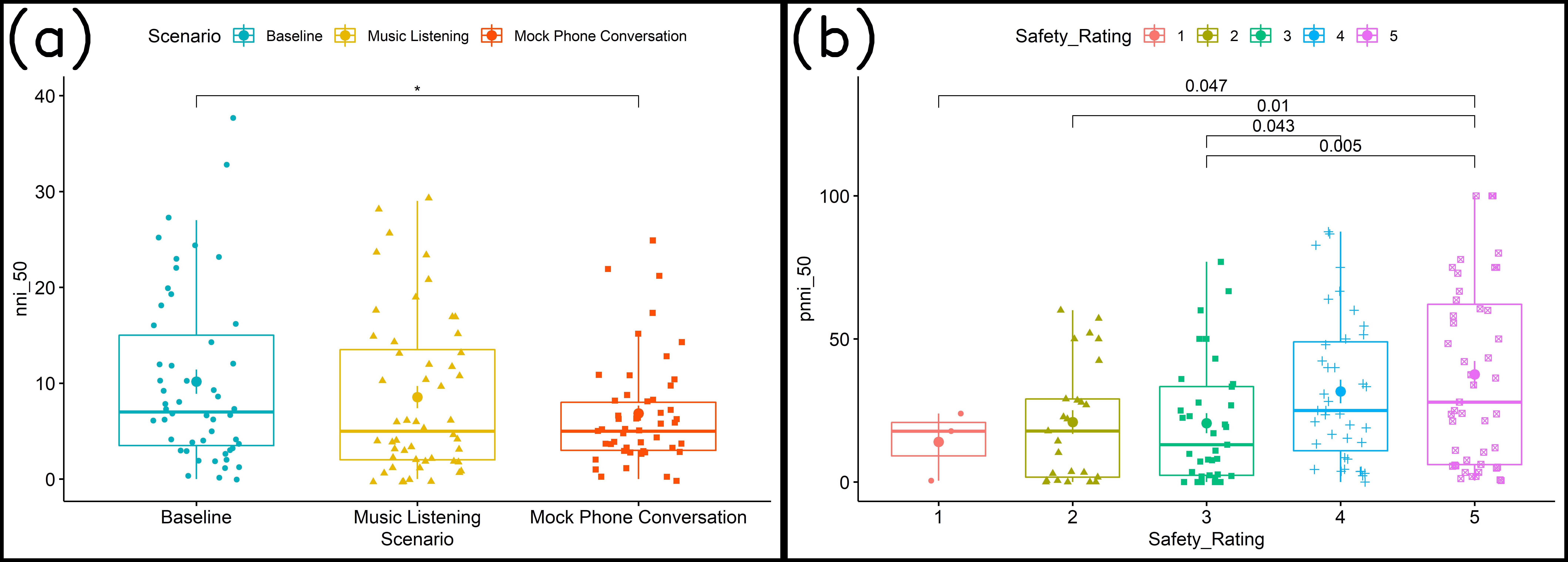}
    \caption{Pnni50-HRV features in (a) different scenarios, (b) cyclists with different levels of sujective safety ratings}
    \label{fig_cognitive:pnni50-HRV}
\end{figure}

Another interesting finding from HRV is in the pnni-50 feature. Cyclists in the mock phone conversation scenario have significantly lower pnni-50 values than the baseline with $\beta = -4.777, SE = 1.785, p = 0.00849$ (Figure \ref{fig_cognitive:pnni50-HRV}a). In addition, pnni-50 is significant with cyclists' subjective safety ratings. As shown in Figure \ref{fig_cognitive:pnni50-HRV}b, cyclists with higher levels of subjective safety ratings also have higher pnni-50 values in general.

\subsubsection{EDA}
\begin{figure} 
    \centering
    \includegraphics[width=\linewidth]{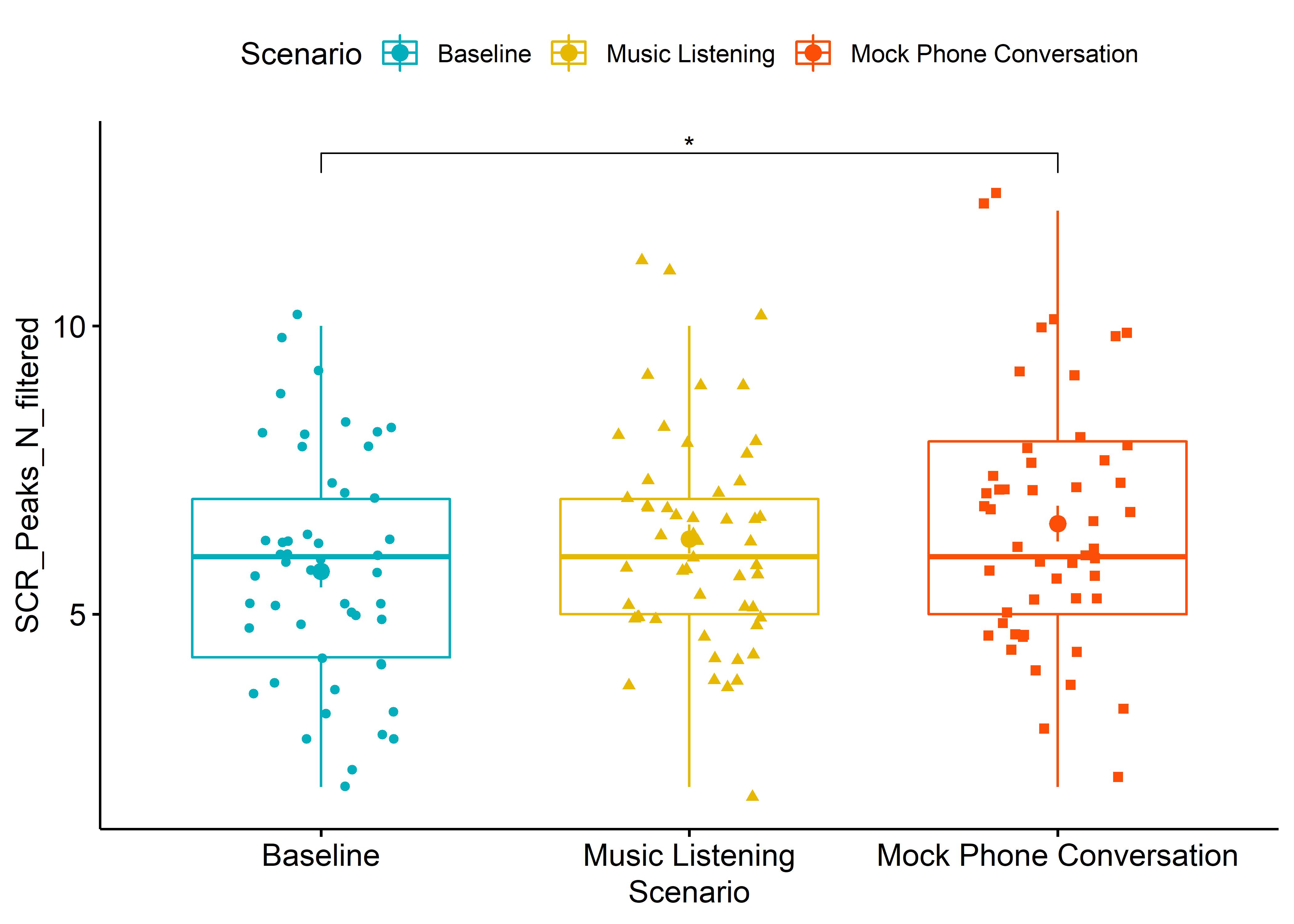}
    \caption{The number of SCR peaks in different scenarios}
    \label{fig_cognitive:SCR}
\end{figure}

The numbers of SCR peaks for the Baseline, Music listening, and Mock phone conversation scenarios are 5,74, 6.30, and 6.57. The Mock phone conversation scenario has a significantly higher number of SCR peaks than the Baseline ($\beta = 0.771, SE = 0.341, p = 0.0258$), as shown in Figure \ref{fig_cognitive:SCR}. The age factor is also found to be significant where the younger group has a lower number of SCR peaks than the older group ($\beta = -0.881, SE = 0.341, p = 0.0258$)

The mean amplitude of SCR peaks for the Baseline, Music listening, and Mock phone conversation scenarios are 0.0707 $\mu S$, 0.0695$\mu S$, and 0.0832 $\mu S$. The Mock phone conversation scenario appears to have a slightly higher mean amplitude of SCR peaks than the Baseline, but the differences are not significant. No other significant differences are found at a 95\% confidence level.

\section{Discussion}
We measured three subjective ratings from the post-experiment survey: safety, comfort, and distraction. For the two types of secondary tasks, not surprisingly, the Mock phone conversation is rated as the most distracting scenario, as it requires both listening to the audio (input) and speaking out the response (output). And for music listening, the cyclists only need to listen to the audio (input). The safety rating is correlated with the distraction rating, lower distraction rating scenarios have higher safety ratings. In terms of comfort rating, no significant results are found between the Baseline and Music listening scenarios, both scenarios have a higher comfort rating than the Mock phone conversation scenario. 

Different levels of cognitive distraction have different effects on cycling behavior and physiological response. For cycling performance, the Music listening scenario has a significantly higher average speed and input power than the Mock phone conversation, as cyclists have a lower subjective rating on the distraction of music listening scenario, they are more confident to keep a higher speed in the IVE with more input power, although the safety rating of music listening is lower than the Baseline.

In a previous virtual reality-based distracted cycling study, it was found that those in a low perceptual load (visual distraction) VR cycled at a higher intensity despite greater pain \cite{wender2022virtual}. An earlier real road study also reported that telephoning coincided with reduced speed, reduced peripheral vision performance, and increased risk and mental effort ratings \cite{de2010mobile}. That study created different levels of perceptual load by displaying different items in the VR for the item detection task. In our study, we generated different levels of cognitive distraction, and with low cognitive load (music listening), we observe a similar effect as a visual distraction. With high cognitive load (mock phone conversation), the adaptive cycling performance includes lowering speed, less input power, and less head movement, indicating a degraded perception ability of the surrounding environment, which is aligned with previous research findings with drivers made fewer saccades, spent more time looking centrally and spent less time looking to the right periphery \cite{harbluk2002impact}. 

However, in terms of lateral lane position, our findings of cycling performance under the influence of cognitive distraction are different from driving. With the findings from our previous experiment \cite{guo2023psycho}, we design bike lanes for the whole road in this experiment, which is different from the real road of the shared bike lane with vehicles. The introduction of bike lanes in the last experiment was found to help the bicyclist to keep closer to the road curbside. In this study, a similar effect is found as there are no significant differences between different scenarios in lateral lane position. While in driving-related studies, cognitive load led to a diminished standard deviation of lateral position, implying a better lane-keeping performance. However, a systematic comparison of time-to-line crossing calculations suggested a degraded safety margin of lane keeping \cite{li2018does}. 

Music listening has been found to be related to emotional arousal, which has the potential to affect cycling performance. For example, listening to preferred music showed no ergogenic benefit during repeated anaerobic cycling sprints when compared to non-preferred music. However, preferred music increased motivation to exercise and decreased perceived exertion \cite{ballmann2019effects}. The cycling task in this experiment is low intensity, listening to preferred music, as indicated by the survey results with higher familiarity and preference of the song played in the Music listening scenario, which may help to explain the increased speed and input power. Cyclists' engagement in the music also leads to a decreased standard deviation of speed, therefore, they will keep a high speed while avoiding any additional speed changes during the whole cycling process. 

For the physiological response, no significant results are found in the mean HR, but the HR change points results showed that there are fewer heart rate change points in the mock phone conversation scenario than in the Baseline and Music listening scenarios. While there are multiple HRV features, research shows that HRV-HF can be used a short-term measure of cognitive load \cite{charles2019measuring}. The HRV-HF is slightly higher in the mock phone conversation compared to the music and baseline condition. Previous studies showed a positive correlation between cognitive workload and the HRV measures including the HRV-HF feature \cite{solhjoo2019heart,charles2019measuring}. This finding suggests that the mock phone interview has a higher cognitive load as compared to the other two scenarios. 

We also observe a slightly lower HRV-pnni50 value for the mock phone interview scenario. Research shows that pnni50 is correlated with the parasympathetic (PNS) activity \cite{shaffer2017overview}. Lower pnni50 and the resulting lower PNS activity during the mock phone interview may show a higher workload level for cyclists during this condition with respect to the other conditions \cite{kim2018stress,delliaux2019mental}. 

The mock phone conversation has a higher number of SCR peaks. Previous research shows that higher skin conductance response is correlated with an increase in cognitive load. Our results indicate that mock phone conversation based on the skin conductance activity shows an increased cognitive load \cite{ayres2021validity,charles2019measuring}. The physiological responses collectively show that these measures can help reliably differentiate between different levels of cognitive load resulting from cognitive distraction during cycling.

Demographic differences are found in several aspects. Generally speaking, participants who hold more positive attitudes toward cycling will go more on the left (closer to the vehicle lane), as they may be more confident about their ability to control the bike. The HR change points data reveals the gender difference as male participants have a significantly lower frequency of increasing HR than female participants. The differences vary in different levels of cognitive load. In the Music listening scenario, the younger group has a significantly higher cycling speed and input power as they are more used to the selected music. 
As prior research has demonstrated that people’s judgments and behaviors are relative, and based on their context \cite{sharif2016effect, sharif2021effect}, the frequency at which people normally engage in a behavior is likely to affect how distracting this behavior is while driving. In the Mock phone conversation scenario, participants who listen to music a lot (>4 hours daily) have a lower standard deviation of speed, indicating they are more engaged in the music. The physiological response also shows that female cyclists are more affected by the cognitive load in the Mock phone conversation scenario, as (1) male participants have a higher head movement variation than female participants, and (2) Female participants have a higher mean amplitude of SCR peaks. These results highlight the groups of people that require more attention when studying cyclist cognitive distraction (young people who listen to music a lot in their daily life and female cyclists under the effect of higher cognitive distraction such as talking on the phone).

\section{Conclusion}

This research explores the effect of different cognitive distractions on bicyclists' physiological and behavioral changes. In an immersive virtual environment, a bicycle simulator with multiple physiological sensing devices is utilized to collect bicyclists’ behavioral and physiological responses on the same road design with bike lanes. Data collection includes demographic information (age, gender, biking attitude), engagement in a secondary task such as music listening and talking on the phone during their daily life or cycling, cycling performance in the simulator (speed, lane position, input power, head movement) and physiological responses (heart rate, skin temperature). Results from 75 participants who rode on a bicycle simulator through a virtual environment indicate that (1) Cyclists would have a significantly higher speed, a lower standard deviation of speed, and higher input power in the music listening scenario; (2) When talking on the phone, cyclists had a lower speed with less input power and less head movement variation; (3) When listening to music, cyclists who had a strong habit of daily music listening ($>$ 4 hours/day) had a higher engagement in the music, with a significantly lower sd of speed. Male cyclists stayed closer to the vehicle lane and had a higher head movement variation; (4) Lane position is not affected by the scenario, this may be the effect of introducing bike lanes in the environment, and (5) Several HRV (HF, pnni-50) and EDA features (numbers of SCR peaks) are sensitive to cyclists' cognitive load changes in the IVE simulator.



\bibliographystyle{IEEEtran}

\bibliography{sample}

\newpage

\begin{IEEEbiography}[{\includegraphics[width=1in,height=1.25in,clip,keepaspectratio]{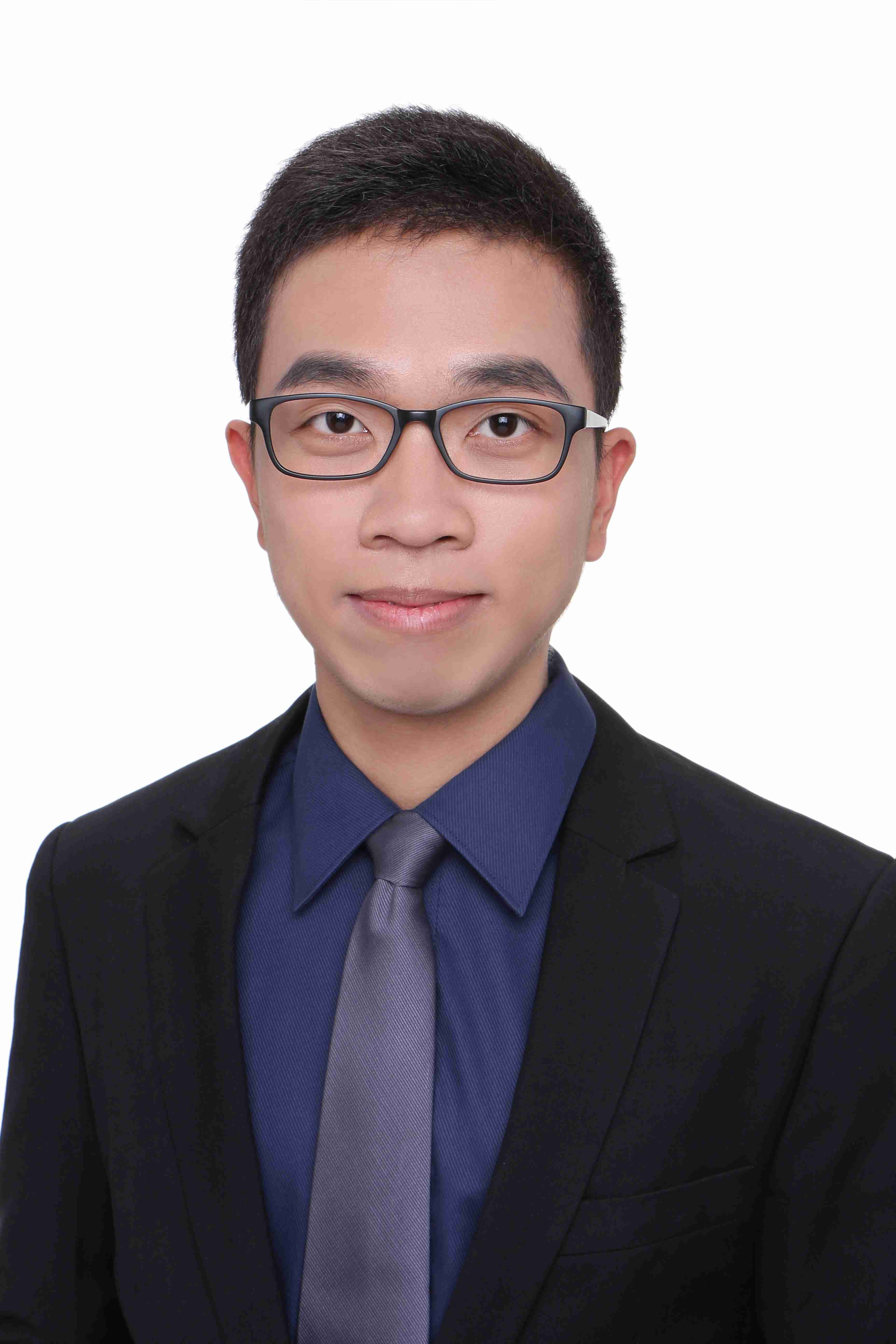}}]{Xiang Guo} Dr. Xiang Guo earned his phd from the Engineering Systems and Environment department at University of Virginia. He received his BSc. and the MSc. degrees in transportation engineering from Beihang University, Beijing, China. His research interests include traffic safety, human factors, human performance modeling, virtual reality and mixed reality.
\end{IEEEbiography}

\begin{IEEEbiography}[{\includegraphics[width=1in,height=1.25in,clip,keepaspectratio]{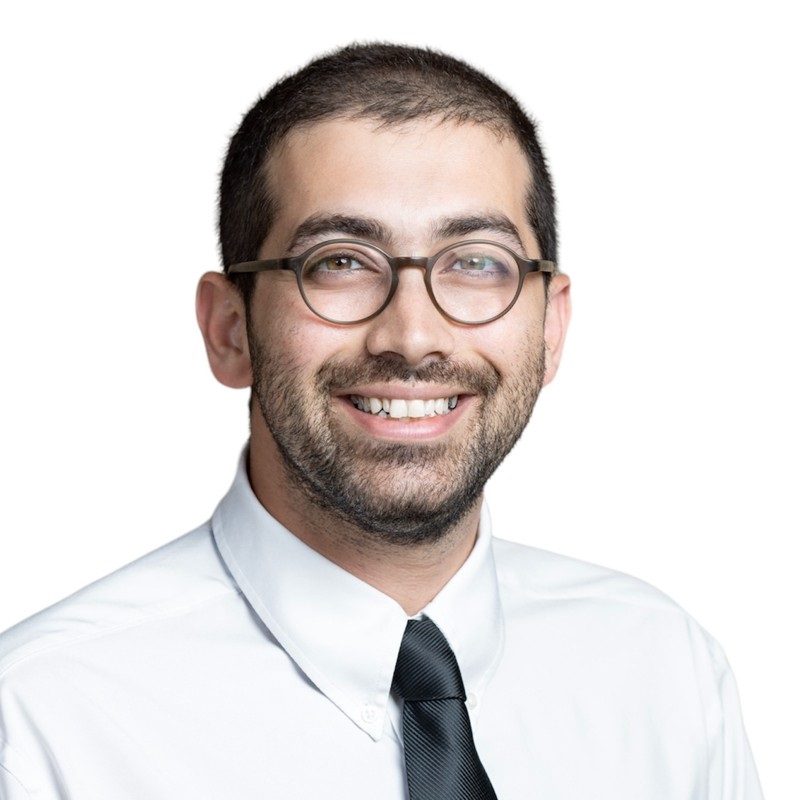}}]{Arash Tavakoli} Dr. Arash Tavakoli is a Postdoctoral Scholar at Stanford University, Department of Civil and Environmental Engineering. He graduated with a Ph.D. in Civil Engineering from the University of Virginia. He has also earned his BSc. and MSc. in Civil Engineering from the Sharif University of Technology and Virginia Tech, respectively. Arash’s research interest lies in the intersection of transportation engineering, computer science, and psychology.
\end{IEEEbiography}

\begin{IEEEbiography}[{\includegraphics[width=1in,height=1.25in,clip,keepaspectratio]{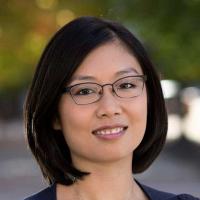}}]{T. Donna Chen} Dr. T. Donna Chen is an Assistant Professor in the Department of Engineering Systems and Environment at the University of Virginia. Her research focuses on sustainable transportation systems (in particular modeling the impacts of new vehicle technologies systems on traveler behavior and the environment), travel demand modeling, transportation economics, and crash safety. Prior to joining academia, Dr. Chen worked in the consulting industry as a transportation planning engineer and has experience with roadway design, cost estimation, and traffic operation analyses.
\end{IEEEbiography}

\begin{IEEEbiography}[{\includegraphics[width=1in,height=1.25in,clip,keepaspectratio]{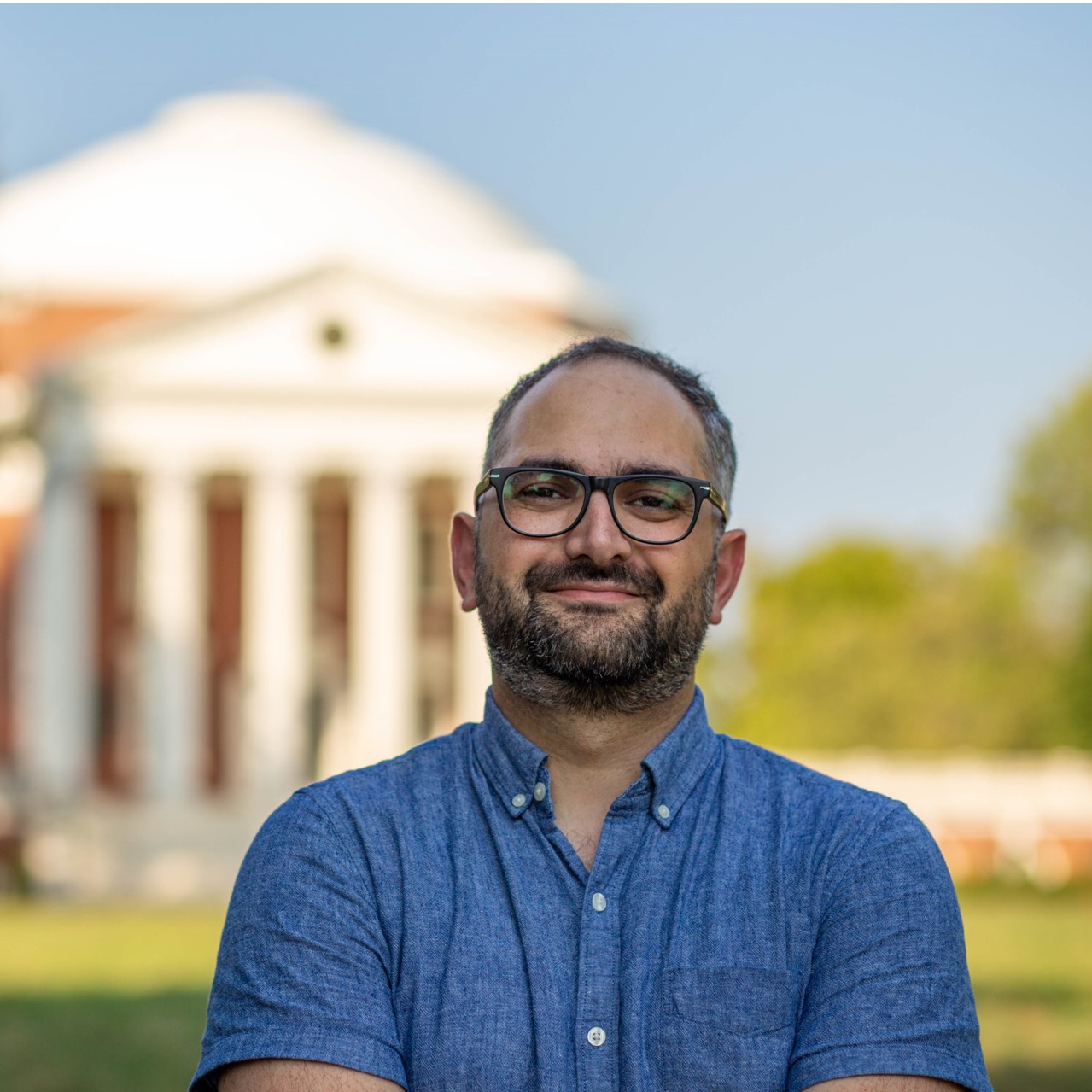}}]{Arsalan Heydarian} Dr. Arsalan Heydarian is an Assistant Professor in the Department of Engineering Systems and Environment as well as the UVA LINK LAB. His research focuses on user-centered design, construction, and operation of intelligent infrastructure with the objective of enhancing sustainability, adaptability, and resilience in future infrastructure systems. Specifically, his research can be divided into four main research streams: (1) intelligent built environments; (2) mobility and infrastructure design; (3) smart transportation; and (4) data-driven mixed reality. Dr. Heydarian received his Ph.D. in Civil Engineering from the University of Southern California (USC), M.Sc in System Engineering from USC, and B.Sc. and M.Sc in Civil Engineering from Virginia Tech. 
\end{IEEEbiography}

 




\vfill

\end{document}